\documentclass[prx,showpacs,showkey,floatfix,onecolumn,preprint]{revtex4-1}
\usepackage{amsmath}
\usepackage{amssymb}
\usepackage{amstext}
\usepackage{amsopn}
\usepackage{amsxtra}
\usepackage{amsfonts}
\usepackage{dcolumn}
\usepackage{graphicx}
\usepackage{graphics}
\usepackage{bm}
\usepackage{hyperref}
\usepackage{color}
\usepackage{ulem}
\pdfoutput=1
\begin{document}
\title{Isotope effect in superconducting n-doped SrTiO$_3$}
\author{A.~Stucky}
\author{G.~Scheerer}
\author{Z.~Ren}
\author{D.~Jaccard}
\author{J.-M.~Poumirol}
\author{C.~Barreteau}
\author{E.~Giannini}
\author{D.~van~der~Marel}\email{Correspondence to dirk.vandermarel@unige.ch}
\affiliation{Department of Quantum Matter Physics, Universit\'{e}
de Gen\`{e}ve, CH-1211 Gen\`{e}ve 4, Switzerland}
\date{\today}
\begin{abstract}
{ We report the influence on the superconducting critical
temperature $T_c$ in doped SrTiO$_3$ of the substitution of the
natural $^{16}$O atoms by the heavier isotope $^{18}$O. We observe
that for a wide range of doping this substitution causes a strong
($\sim 50 \%$) enhancement of  $T_c$. Also the magnetic critical
field $H_{c2}$ is increased by a factor $\sim 2$. Such a strong
impact on $T_c$ and $H_{c2}$, with a sign opposite to conventional
superconductors, is unprecedented. The observed effect could be
the consequence of strong coupling of the doped electrons to
lattice vibrations (phonons), a notion which finds support in
numerous optical and photo-emission studies. The unusually large
size of the observed isotope effect supports a recent model for
superconductivity in these materials based on strong coupling
to the ferroelectric soft modes of SrTiO$_{3}$.}
\end{abstract}

\maketitle
SrTiO$_{3}$ is a para-electric insulator which becomes
ferroelectric when 35$\%$ or more of the oxygen is substituted
with the isotope $^{18}$O \cite{mueller1979,itoh1999,rowley2014}.
Due to electron-phonon coupling doped charge carriers form a
polaronic liquid at small
concentration~\cite{eagles1969,calvani1993,gervais1993,vanmechelen2008,meevasana2010,vandermarel2011,boschker2015,chen2015,wang2016,cancellieri2016}
and the material becomes  superconducting with a doping dependent
critical temperature ($T_c$) below $\sim1$ Kelvin both for
bulk~\cite{schooley1964,koonce1967,pfeiffer1969,bednorz1988,suzuki1996,lin2014}
and interfaces~\cite{reyren2007,caviglia2008}.
Several aspects of this material
are not usually encountered in conventional superconducting
metals. These include a multi-valley bandstructure
\cite{cohen1964} (however presently known not to apply to
SrTiO$_{3}$), the structural transition at around 100
K\cite{appel1969}, and the low density of charge
carriers\cite{eagles1969,zinamon1970,jarlborg2000,klimin2012,gorkov2016}.
More recently a connection between the near ferroelectric
instability and the superconductivity of SrTiO$_{3}$ has been
conjectured\cite{rowley2014,edge2015}.

Here we report on the superconducting properties of doped
SrTiO$_{3-y}$ with partial $^{18}$O/$^{16}$O isotope substitution,
and observe that substituting 35$\%$ of the heavier $^{18}$O for
$^{16}$O increases $T_c$ by a factor of approximately $1.5$. The
sign of the observed isotope effect is opposite to one in conventional superconductors and the magnitude much
stronger. The
unusual size and sign of the isotope effect may be caused
by the near ferroelectric
instability, the polaronic nature of the charge carriers, or a
combination of those.

Undoped SrTiO$_3$ crystals were annealed under pure $^{18}$O$_2$
atmosphere. During this procedure, part of the $^{16}$O atoms in
the structure is substituted by $^{18}$O atoms. The isotope
substitution ratio has been quantified by three independent and
complementary experimental techniques. We estimate the amount of
isotope substitution to be $x(^{18}O)/x(^{16}O)\simeq0.35$ (see
{\it Methods}). N-type doped samples were prepared by subjecting
the samples to a reduction process. Three samples were made with
charge carrier density of n=0.004\,$nm^{-3}$, n=0.02\,$nm^{-3}$
and n=0.07\,$nm^{-3}$ representative of a large portion of the
phase diagram. At each doping level of SrTi$^{18}$O$_{3-y}$, one
parent SrTi$^{16}$O$_{3-y}$ sample was annealed under the same
conditions and at the same time, thus providing a reference for
the isotope effect on the superconducting properties. Details of
preparation and characterization can be found in {\it Methods}.
The samples were cooled down to 25~mK in a $^{3}$He-dilution
refrigerator. A current between 10 and 50\,mA (corresponding to a
density between 1.6 and 8\,A/cm$^{2}$) was flowing in the sample.
The longitudinal voltage was acquired to probe the intrinsic
resistivity; the resulting $\rho(T)$ curves are reported in
Fig.~\ref{fig1}~(a-c). The "onset" transition temperatures are
extracted from this data as the crossing point between the linear
extrapolation of the normal state resistivity above $T_c$ and the
linear extrapolation of the resistivity drop at the transition.
The error bar is estimated as the
temperature range over which the derivative $d \rho(T)/d
T$~changes. Fig.~\ref{fig1}~(e) shows the extracted transition
temperatures as a function of the charge carrier density for all
samples. The values of $T_c$ indicate a systematic increase of the
superconducting transition temperature with the presence of the
heavier isotope. Figure~\ref{fig1}(d) shows the magnetic
AC-susceptibility for the two highly doped samples. This was
measured in a two-coil set-up whose pick-up coil voltage change
was amplified by a standard lock-in (feeding current 0.05 mA at a
frequency of 977 Hz, providing a magnetic field of 0.014
mT)\cite{jaccard2010}. The transition temperatures, obtained from
transport (panel (c)) and AC-susceptibility measurements (panel
(d)), well agree with each other.

The BCS weak coupling limit gives an isotope coefficient
$\alpha=-d(\ln T_c) / d(\ln M_{})=0.5$ (where $M_{}$ is the oxygen
isotope mass), corresponding to a $T_c$ shift of $-5\%$ as opposed
to approximately $+50\%$ in the data presented here. The first
main observation is therefore that, contrary to most
superconducting materials, the isotope coefficient in SrTiO$_{3}$
is negative. In some other rare cases $\alpha_{} < 0$ has been
observed: the pure uranium ($\alpha= -2.2$) \cite{fowler1967}, the
high-$T_c$ superconductor Bi$_2$Sr$_2$Ca$_2$Cu$_3$O$_{10}$
($\alpha=-0.1$) \cite{bornemann1991}, and the metal hydride PdH(D)$_{x}$
\cite{stritzker1972,miller1974} $(-0.3<\alpha<-0.1)$.
Controversial sign changes of the isotope coefficient have been
observed in (Ba,K)Fe$_{2}$As$_{2}$ ($\alpha_{}= -0.2$)
\cite{shirage2009,liu2009} (due to differences in the sample
composition), and in pure lithium under high pressure ($\alpha$
changes with increasing pressure ~\cite{schaeffer2015}). We will
not dwell on the physical origins of the isotope effect in these
cases, which are certainly different in the case of uranium, and
possibly different in the other examples as well. Our second main
observation is, that also the magnitude of the isotope effect is
remarkable: an overall enhancement of $T_c$ of a factor $1.5$ is
observed at all doping levels, which corresponds to a negative and
large coefficient $\alpha_{} \sim -10$. This is possibly the
strongest isotope enhancement of $T_c$ observed in any material so
far.

From the measurement of the electrical resistivity at the
superconducting transition in magnetic field, we could estimate
the upper critical field $H_{c2}(T)$ and the effect that the
isotope substitution can have on it. These results are summarized
in Fig.~\ref{fig2}. For each sample we plot the onset of
$\rho(T,H)$, defined as the crossing point of the linear
extrapolations of the normal state resistance and the slope of the
resistance at the transition. The $H_{c2}(T)$ line of
$^{18}$O-substituted SrTiO$_{3}$ is far above that of the pristine
sample, at each doping level. The isotope effect does not only
enhance the critical temperature, but strengthens the
superconductivity in a magnetic field as well, up to a maximum
$\mu_{0}H_{c2}(T=0)\simeq 0.25$ tesla at optimal doping. For all dopings the observed isotope effect on $H_{c2}$  corresponds to $\beta =-d\ln H_{c2}/d \ln M \approx -19$.
We notice that the samples with lower charge carrier density exhibit a higher
critical field, despite of the lower critical temperature. This is
not surprising and it is expected on the basis of the band
structure reported in [\onlinecite{vandermarel2011}] and the
two-gap superconductivity reported
for SrTi$_{1-x}$Nb$_x$O$_{3}$ [\onlinecite{binnig1980}](since gap anisotropy is in general affected by impurity scattering, this may {\it a priori} depend on the kind of doping, {\it e.g.} oxygen depletion, Nb substitution or La substitution). As a
matter of fact, the different doping levels shown here correspond
to the occupation of different bands \cite{vandermarel2011}, which
contribute differently to the superconducting pairing. The
temperature dependence of the critical field does not change even
when the critical temperature and field are enhanced by the
isotope substitution.

We measured the resistivity under a magnetic field in both
orientations perpendicular to the current (with the current always in the
same direction in the $ab$-plane of the crystal). Both
measurements provide the same result, thus ruling out any possible
surface effect on the superconducting behavior. The lowermost
panel of Fig.~\ref{fig2} shows the two identical $H_{c2}(T)$
transition lines measured in the two different field orientations.

There are three infrared active
modes in SrTiO$_{3}$, having transverse frequencies 67, 21 and 2
meV and Fr\"ohlich coupling constants $\alpha_{ep}=1.6$, 0.45 and
0.02, respectively (see Table 1 of
Ref.~\onlinecite{devreese2010}). The (Fr\"ohlich type)
electron-phonon coupling in SrTiO$_3$ is dominated by coupling to
the optical phonons at 67 and 21
meV~\cite{devreese2010,meevasana2010}; these phonons are to a
large extend responsible for the twofold mass-enhancement of the
charge carriers observed with
optics\cite{vanmechelen2008,devreese2010}. Note that the coupling
constant of the soft ferroelectric mode at 2 meV can not be
correctly estimated from the Fr\"ohlich mode; it is certainly
larger, among other things due to the strong anharmonicity of this
mode\cite{edge2015,kedem2016}. The doping dependence of the
effective mass  in units of the bare band-mass is described by the
phenomenological expression \cite{vandermarel2011} $m^*/m= 2.0+1.2
\exp{(-n/n_0)}$, where $n_0=0.084$ nm$^{-3}$ is an empirical
factor. For all doping concentrations where superconductivity is
observed the quasiparticles are slow compared to the vibrational
degrees of freedom~\cite{vandermarel2011}, which corresponds to
the anti-adiabatic limit. The quasiparticles in this limit are
commonly referred to as ``polarons''. In case the interactions are
not too strong the polarons form a Fermi-liquid.
Alexandrov\cite{alexandrov1992} has pointed out in this context
that the critical temperature of the superconducting transition is
determined by a BCS-like formula, but with the renormalized
density of states replacing the bare one. In conventional models
of electron-phonon coupling the effective coupling constant
describing the pairing interaction, $\lambda$, is invariant under
isotope substitution. (We use a single parameter $\lambda$ to
describe the combined phonon mediated and Coulomb interaction.)
However, this does not take into account aforementioned
band-renormalization, due to which it should be replaced with the
effective parameter
\begin{equation}\label{lambdastar}
\lambda^*=\lambda\frac{m^*}{m}
\end{equation}
The corresponding expressions for the isotope coefficient are
detailed in Eqs. \ref{polaron2} and \ref{polaron3} of the Methods
section. The parameters in Table \ref{table_theory} imply that $\alpha_{}\sim-1.7\pm 0.3$, which is larger
than the BCS result and of opposite sign.
For the isotope coefficient of the
upper critical field the model gives $\beta_{}=-d(\ln
H_{c2})/d(\ln M) \sim-3.9\pm 0.5$.
\begin{table}[ht]
\begin{tabular}{|c|c|c|c|c|c|c|c|c|c|}
  \hline
n               &$\epsilon_{F}^*$ &  $\lambda^*$ & $\omega_o/\epsilon_F^*$& $d\ln m^* / d \ln M$ & $\alpha_{}$& $\beta_{}$\\
nm$^{-3}$ &meV                &                          &                                &                                     &              &\\
  \hline
$0.004$     &2.9                  &  0.091           &  17                          & 0.25                            & -2.0 & -4.5\\
$0.020$     &7.1                  & 0.104           &  7                          & 0.24                            & -1.8 & -4.1\\
$0.100$     &20                   & 0.115           &  2.5                          & 0.20                            & -1.3 & -3.0\\
  \hline
\end{tabular}
\caption{Parameters for doped SrTiO$_3$.
For the fourth column $\hbar\omega_0=50$meV was used.
See Ref.~\onlinecite{vandermarel2011} for the values of $\epsilon_F$ as a function of carrier concentration. The the log-derivatives of the mass and the $\alpha_{}$- and $\beta_{}$-coeffients and were calculated with Eqs.~\ref{polaron3}, ~\ref{polaron2} and ~\ref{hc2b}  (Methods) respectively. }\label{table_theory}
\end{table}

Hence the unusual isotope effect that we observe in the
experiments is in principle not unexpected given the polaronic
nature of the charge carriers. Note that $T_c$ is a very sensitive
function of $ \lambda^*$ due to the fact that in STO this
parameter (and consequently also $T_c$) is very {\it small}.
The situation is quite different
in this respect from that in cuprate high temperature
superconductors, for which the $^{18}$O isotope substitution
causes $T_c$ to become smaller\cite{khasanov2008,bussmann2012}.
The fact that in SrTiO$_3$ the experimental isotope coefficients
$\alpha_{}\sim -10$ and $\beta\sim
-19$ are still considerably larger than expected from
aforementioned band-renormalization, calls for deeper theoretical
analysis of the electronic structure and the effective
polaron-polaron interactions in these materials. A possible clue
comes from the recent theoretical study by Edge {\it et al.}
\cite{edge2015}, whom postulated that pairing of electrons in
SrTiO$_{3}$ is driven by the ferroelectric soft mode fluctuations
in the proximity of a quantum critical phase transition. Following
Kedem {\it et al.}\cite{kedem2016} the coupling parameter close to
the ferroelectric quantum critical point is
\begin{equation}\label{qcp}
\lambda \propto \frac{\alpha_{ep}^2}{\omega_s}\frac{M_{}^{z\nu}}{\left|M_{c}-M_{}\right|^{z\nu}}
\end{equation}
where $\alpha_{ep}$ is the electron-phonon coupling constant,
$\omega_s$ a vibrational energy scale, $M_{c}$ the mass at the
critical point and $z\nu$ the critical exponent of the system
having the mean-field value $z\nu=0.5$. For harmonic modes
$\alpha_{ep}^2\propto M^{-1/2}$, which leads to a mass independent
$\lambda=\alpha_{ep}^2/\omega_s$ since also $\omega_s\propto
M^{-1/2}$. Following the description of Edge {\it et al.} the
effect of doping is to increase $M_{c}$, which for the relevant
doping range ($>10^{19}$ cm$^{-3}$) falls above the mass of
$^{18}O$. Consequently at these doping levels there is no
ferroelectric instability (corresponding to a zero in the
denominator) for any (partial or complete) $^{18}O$ isotope
substitution. Yet,  according to Eq. \eqref{qcp}, even far from
the ferroelectric instability $\lambda$ is a sensitive function of
the isotope mass. Based on these considerations Edge {\it et al.}
predicted a 30 \% increase of $T_c$ for 35\% isotope substituted
samples. Note that the renormalization of $\lambda$ due to
polaronic band narrowing occurs regardless of the pairing
mechanism, hence Eqs. \eqref{lambdastar} and \eqref{qcp} need to
be combined. Taken together these two effects do indeed account
-with a large margin- for the observed value, $\alpha_{}\sim -10$,
of the isotope coefficient.

We have studied the critical temperature $T_c$ and the upper
critical field $H_{c2}$ of partially (35~\%) $^{18}$O isotope
substituted doped SrTiO$_3$. We observe a strong (factor $\sim
1.5$) enhancement of $T_c$ and (factor $\sim 2$) of $H_{c2}$. Both
the sign and the size of these two effects are unusual, and
indicate that the superconductivity is not of the conventional BCS
variety. The effect may be the consequence of a combination of two
effects: polaronic band-narrowing and coupling to soft phonons
responsible for the ferroelectric instability in these materials.

\vspace{5mm}
\noindent{\bf Acknowledgements}

\noindent
We thank Prof.
Torsten Vennemann (University of Lausanne) for the sample characterization using mass-spectroscopy.
We are grateful to Alexander Balatsky, Nicola Spaldin, Hugo Keller, Daniel Khomskii and Jose Lorenzana for illuminating discussions.
This work was supported by the Swiss National Science Foundation through grants no.
$200021-153405$, $200021-162628$ and through the National Center of Competence in Research (NCCR) MARVEL.





\newpage
\noindent{\large {\bf Methods}}\label{methods}

\vspace{5mm}
\noindent{\bf Sample processing}

\noindent For the present study, we chose SrTiO$_{3}$ samples from
various sources and different morphologies: commercial crystals of
pure SrTiO$_{3}$, commonly purchased as substrates for thin film
deposition by MTI Corp.; superconducting crystals of Nb-doped
SrTiO$_{3}$, with a Nb:Ti nominal ratio ranging from 0.001 to
0.05, from CrysTech; home-processed ceramic samples of SrTiO$_{3}$
obtained from binary reaction between the binary oxide components.
The isotope substitution of $^{18}$O for $^{1}$$^{6}$O in
crystalline samples was achieved following a procedure similar to
that reported by Itoh {\it et al.} \cite{itoh1999}. We carried out
a three step cycle where in each step the crystals were
re-annealed in a new sealed quartz reactor filled with pure
$^{18}$O$_{2}$. Based on the amount of oxygen in the sample
(between $5\cdot10^{-5}$ and $6.5\cdot10^{-5}\,mol$), and  the
volume of the quartz tubes (between $12$ and $15 \,cm^{3}$), and
assuming that equilibrium has been reached at the end of each of
the 3 consecutive annealing steps, the expected value of $^{18}$O
substitution would be between $99.4$ and $99.9\%$. As a matter of
fact, the procedure followed for substituting the isotope $^{18}$O
for $^{16}$O was only partially effective and did not allow full
isotope substitution. The actual amount of oxygen isotope
substitution proved to be $\sim$35~\%, as directly measured by
three independent and complementary experimental techniques
described in the following.

In the ceramic samples the isotope substitution was obtained by
first oxidizing the pure metals Sr and Ti in pure $^{18}$O$_{2}$
atmosphere, then reacting Sr$^{18}$O and Ti$^{18}$O$_{2}$ to form
SrTi$^{18}$O$_{3}$.
Processing of ceramic samples was highly
time-consuming and costly, and yielded only a small amount of
SrTi$^{18}$O$_{3}$; however, this was a crucial reference for the
calibration and optimization of the $^{18}$O substitution in the
crystalline samples.

The isotope-substituted samples from the various sources were then
subjected to reduction treatments (10$^{-7}$\,mbar at
1000$^\circ$-1350$^\circ$C) in order to create oxygen vacancies
and tune the doping of charge carriers.
For each doping level of
each $^{18}$O-substituted sample, a twin $^{1}$$^{6}$O-sample
(either crystal or ceramic) was subjected to the same annealing
conditions and reduction treatments, at the same time and in the
same furnace.
By virtue of this twin-treatment a complete range of
SrTi$^{16}$O$_{3}$/SrTi$^{18}$O$_{3}$ twin samples was obtained.
This is the optimal procedure to limit ambiguities due to sample
processing when comparing the  physical properties of the
$^{16}$O- and $^{18}$O-samples.
For the study of the
superconducting behavior of SrTi$^{18}$O$_{3}$ through transport
measurements in a dilution fridge, only oxygen-reduced single
crystals of SrTiO$_{3}$ were used.
These samples, as well as their
processing conditions are reported in Table\,\ref{table_samples}.

\begin{table}[ht]
\begin{tabular}{|c|c|c|c|c|r|} \hline
Sample & oxygen & Temp & time & n  & RRR \\
 & isotope & $^\circ C$ & $hours$ & $nm^{-3}$ & \\ \hline
MTI-1 & $^{16}$O$_{2}$ & 1050 & 36 & 0.0041 & 1383 \\
MTI-2 & $^{16}$O$_{2}$ & 1200 & 36 & 0.020  & 389 \\
MTI-3 & $^{16}$O$_{2}$ & 1350 & 36 & 0.070 & 106 \\
MTI-4 & $^{18}$O$_{2}$ & 1050 & 36 & 0.0044  & 1000 \\
MTI-5 & $^{18}$O$_{2}$ & 1200 & 36 & 0.012  & 388 \\
MTI-6 & $^{18}$O$_{2}$ & 1350 & 36 & 0.15  & 45 \\ \hline
\end{tabular}
\caption{Sample identifier, oxygen isotope, annealing temperature,
annealing time, charge carrier density measured from the Hall
constant and residual resistance ratio (RRR=R(300\,K)/(R(4\,K)) of
the samples reported in this manuscript.\label{table_samples}}
\end{table}

\vspace{5mm}
\noindent{\bf Determination of the isotope substitution level}

\noindent The amount of $^{18}$O/$^{16}$O substitution is commonly estimated
from the mass uptake of the sample \cite{itoh1999}.
However, such
an estimation is not free from uncertainties and ambiguities.
We
want to know how much $^{18}$O actually substitutes for $^{16}$O
under the processing conditions used, and check wether the
response of the material is that expected for the
isotope-substituted $^{18}$O-oxide.
We pursued this goal by three different techniques: mass
spectroscopy, thermogravimetry and infrared optical spectroscopy.
Mass spectrometry was performed on selected pieces of crystals and ceramic samples in the Stable Isotope Laboratory at the University of Lausanne.

These measurements confirmed that oxygen has been substituted throughout the bulk of
the material, and indicated that the amount of $^{18}$O in the
(ceramic and crystalline) samples is about  35~\%.
The results of
this analysis are summarized in Table\,\ref{table_mass_spectro}.
The increment of the substitution of $^{18}$O from 10\% to
$\sim$35~\% between the first and the last step of the treatment is
shown as well.
Such an incomplete isotope substitution is unexpected, having followed a procedure previously reported to be
effective for full substitution \cite{itoh1999}.
We will return to this point below.

\begin{table}[ht]
\begin{tabular}{|l|c|c|}
  \hline
 & &  \\
samples&$\%$~of~$^{16}O$ &$\%$~of~$^{18}O$\\
 & &  \\
  \hline
MTI-0~$^{16}O$& 99.79 & 0.21 \\
MTI-0~$^{18}O$ & 91.81 & 8.19 \\
MTI-1 & 99.72 & 0.28 \\
MTI-4 & 61.18 & 38.82 \\
MTI-2 & 99.68 &0.32  \\
MTI-5 & 65.21 & 34.79 \\
ceramic ~$^{16}O$& 99.35 & 0.65  \\
ceramic ~$^{18}O$ & 65.90 & 34.10 \\
  \hline
\end{tabular}
\caption{Isotope content in various samples as obtained from mass
spectroscopy } \label{table_mass_spectro}
\end{table}

In the thermogravimetric experiment, two crystals of
SrTi$^{18}$O$_{3}$ and SrTi$^{16}$O$_{3}$ were simultaneously
heated in the symmetric furnace of a Setaram TAG24 thermal
analyzer under a flux of pure $^{16}$O$_{2}$.
The same thermal
treatment as used for the substitution process was reproduced in
the thermal analyzer.
The mass loss of the former with respect to
the latter corresponds to the loss of $^{18}$O, replaced back by
the $^{16}$O during this treatment.
The drift of the thermobalance
as a function of time and temperature was measured from a similar
thermogravimetric experiment in which two identical
SrTi$^{16}$O$_{3}$ crystals were used as the sample and the
reference, respectively.
The mass loss associated to the loss of
$^{18}$O is plotted in Fig.~\ref{fig3}~(a).
The total amount of
$^{18}$O substituted by $^{16}$O is found to be 35\%, in good
agreement with the mass spectrometry.

The isotope substitution is expected to affect the phonon modes,
whose frequency shift can be measured by either infrared (IR) or
Raman spectroscopy.
If the reduced mass of an IR vibrational mode
is that of the oxygen atoms, and assuming that everything else
(lattice parameter, atomic positions, zero-point fluctuations)
remains the same upon isotope substitution, the softening for a
complete substitution of $^{16}$O by $^{18}$O would be given by
$\omega(^{18}$O)/$\omega(^{16}$O)$ =\sqrt{16/18} $, which
corresponds to a red-shift of about 6\%.
The IR reflectivity of
the pristine and substituted samples, measured in a Fourier
transform infrared spectrometer, is shown in Fig.~\ref{fig3}~(a)
over the energy range from 350 to 600~cm$^{-1}$ (12.4 to
86.8~meV).
The strong absorption at 480~cm$^{-1}$ (59.5~meV) is
associated to the TO4 phonon mode due to the axe zone-center
displacement of the oxygen octahedra \cite{hlinka2006} and is
strongly dependent on the oxygen isotope mass.
Fig.~\ref{fig3}~(b)
shows the softening of that mode in samples exposed to repeated
thermal treatments under $^{18}$O atmosphere, thus proving that
the isotope substitution has occurred.
The measured shift is a
factor three lower than the $\sim$6\% expected for a complete
isotope substitution, confirming that the procedure followed for
isotope substitution is effectively replacing only $\sim$35\% of
$^{16}$O by $^{18}$O.

The red-shift of the phonon mode is found to be of the same
magnitude in ceramic samples as in crystals, the former having
being processed by oxidation of metal elements in pure
$^{18}$O$_{2}$, and confirms well the amount of substitution
measured by the mass spectrometry and thermogravimetry. One would
expect complete isotope substitution in ceramic samples.
However, the reaction treatment to form SrTi$^{18}$O$_{3}$ from
Sr$^{18}$O and Ti$^{18}$O$_{2}$ is done in SiO$_{2}$reactors at
high temperature, thus bringing it down to the same high
temperature environment as used for crystalline samples.
This leads us to suspect that part of the oxygen of the quartz tubes participates in the equilibrium.
This could be a surface layer, or a small fraction of less stable bonded oxygen.

\vspace{5mm}
\noindent{\bf Electron doping of SrTiO$_{3}$}

\noindent After the $^{18}$O-substitution, n-type charge carriers were
introduced by creating oxygen vacancies through vacuum annealing.
This is a widely used procedure, long known as being successful in
tuning the carrier density and the conductivity of SrTiO$_{3-y}$.
\cite{koonce1967,spinelli2010}.
By optimizing the
annealing temperature between 800$^\circ$C and 1400$^\circ$C and
the annealing time between 20 and 36\,h, we could span a wide
range of charge doping, from $5\cdot 10^{16}$ to $2\cdot 10^{20}$
cm$^{-3}$.
This doping range corresponds to the
underdoped side of the superconducting phase diagram up to the
maximum critical temperature of $\sim$300~mK at $n=10^{20}$
cm$^{-3}$.
The carrier density is obtained by measuring the Hall
effect in a 5-probe configuration using a Quantum Design PPMS
apparatus.
The homogeneity of oxygen depletion is enhanced after
long annealing time.
According to the Hall characterization of a
variety of samples subjected to different reducing treatments, we
have selected, for this study, three pairs of crystalline samples
(three with $^{16}$O ad three with $^{18}$O) with the same oxygen
reductions, having been treated for 36h at 1050$^\circ$,
1200$^\circ$, and 1350$^\circ$C.
The charge carrier density
(measured at 4\,K) in oxygen reduced samples was $2\cdot10^{18}$
cm$^{-3}$, $4\cdot10^{19}$~cm$^{-3}$, and $10^{20}$~cm$^{-3}$, for
the three reduction temperatures, respectively.

According to the literature \cite{koonce1967}, these carrier
densities are expected to correspond to superconducting critical
temperatures of 100~mK, 200~mK and 300~mK, respectively (see
Fig.~\ref{fig1}~(e)).
The electrical resistivity as a function of
temperature of various samples is displayed in
Fig.~\ref{fig1}~(a-c), whereas the magnetic transition in the
AC-susceptibility is shown in Fig.~\ref{fig1}~(d).
The Nb-substituted crystals could not be used for the same study.
The
$^{18}$O-substitution treatment actually affect the Nb doping,
thus modifying the charge carrier density in an uncontrolled way.
Because of the impossibility
to tune and control independently the isotope substitution and the
charge carrier doping in Nb-doped SrTiO$_{3}$, we selected only
O-reduced crystals for this study.

\vspace{5mm}
{\noindent{\bf Back-substitution}

\noindent The two samples treated  under flux of $^{16}$O$_2$ for
the characterization by thermogravimetry (MTI-1 and MTI-4) are
used to study the physical properties after back-substitution.
Fig.~\ref{fig4} shows the infrared reflectivity for the two
back-substituted samples. No shift has been observed in the phonon
frequency  of back-substituted $^{16}$O and back-substituted
$^{18}$O compared to the pristine sample. This confirms the fact
that both samples are indeed completely $^{16}$O back-substituted.
The same doping procedure by oxygen reduction as the one for
samples MTI-1 and MTI-4 (see Table~\ref{table_samples}) has been
applied to the back-substituted samples. Electric transport
measurements down to 30mK and Hall characterization were performed
in parallel on the two samples using the same procedure as
presented in Fig.~\ref{fig1}~(a-c). The results, obtained
following the same way to extract the critical temperatures and to
calculate the charge carrier densities, are compared to the
substituted samples in the Fig.~\ref{fig1}~(e). The general rise
of $T_c$ of the back-substituted samples indicates that repeated
annealing improves the quality of the materials. Despite
performing the annealing of the two back-substituted samples
simultaneously in the same furnace, their electron concentration
is different. This is explained by the fact that, due to the
unavoidable step of balance calibration in the thermo-gravimetry
set-up, the back-substitution process has been repeated twice for
the $^{16}$O sample. Thereby, the two back-substituted samples
didn't have the same level of oxygen vacancies before the
reduction. Nevertheless the crucial observation is that, after the
back-substitution, the superconducting transition temperature of
the back-substituted samples has become nearly the same. This
allows us to conclude that the $T_c$ in SrTiO$_3$ is tuned by the
mass of oxygen isotope. }

\vspace{5mm}
{\noindent{\bf Isotope effect on $T_c$, $H_{c2}$, $\upsilon_F$ and $\lambda$ due to polaronic band-narrowing}

\noindent Due to the anti-adiabatic conditions, the expression for $T_c$ in the relevant doping range reads\cite{vandermarel2011}
\begin{eqnarray}\label{polaron}
k_BT_c&\simeq&0.61\epsilon_F^* \exp{\left\{\sqrt{\frac{\omega_{o}}{\epsilon_F^*}}-\frac{1}{\lambda^*} \right\}}
\end{eqnarray}
with $\epsilon_F^*=\epsilon_Fm/m^*$ the Fermi temperature of the polaron Fermi-liquid.
Since~$\omega_{o}^2 \propto 1/M_{}$, the oxygen isotope coefficient for $T_c$ is
\begin{eqnarray}\label{polaron2}
\alpha_{}&=&-\frac{d\ln T_{c} }{ d\ln{M_{}}}=
\frac{1}{4}\sqrt{\frac{\omega_{o}}{\epsilon_F^*}}+\left\{1-\frac{1}{\lambda^*} -\frac{1}{2}\sqrt{\frac{\omega_{o}}{\epsilon_F^*}}\right\}\frac{d\ln{m^*}}{d\ln{M_{}}}
\end{eqnarray}
The upper critical field in the
clean limit is
\begin{eqnarray}\label{hc2}
H_{c2}&=&\frac{\Phi_0}{2\pi \xi^2} \propto \left( T_cm^*\right)^2
\end{eqnarray}
from which we readily obtain for the isotope coefficient of the upper critical field
\begin{eqnarray}\label{hc2b}
\beta&=&-\frac{d\ln H_{c2} }{ d\ln{M_{}}}=
2\alpha - 2 \frac{d\ln{m^*}}{d\ln{M_{}}}
\end{eqnarray}
Another parameter showing isotope effect is the Fermi-velocity,
for which
\begin{eqnarray}\label{vF}
\frac{\upsilon_{F}^*}{\upsilon_{F}}&=&\frac{m}{m^*} \Rightarrow
\frac{d\ln \upsilon_{F}^* }{
d\ln{M_{}}}=-\frac{d\ln{m^*}}{d\ln{M_{}}}
\end{eqnarray}
This can in principle be measured with high resolution angular
resolved photo-emission. The isotope effect of the London penetration
depth is described by
\begin{eqnarray}\label{lambda}
\frac{\lambda_L^*}{\lambda_L}&=&\left(\frac{m^*}{m}\right)^{1/2} \Rightarrow   \frac{d\ln \lambda_L^* }{ d\ln{M_{}}}=\frac{1}{2}\frac{d\ln{m^*}}{d\ln{M_{}}}
\end{eqnarray}
which can in principle be measured using $\mu$SR.

For SrTiO$_3$ the polaron mass renormalization $m^*/m$ is of intermediate strength~\cite{vanmechelen2008} for which~$m^*/m= 1+\alpha_{ep}/6+\alpha_{ep}^2/40+...$, where $\alpha_{ep}$ is the electron-phonon coupling strength causing the mass-renormalization and the expansion presupposes weak coupling ({\it i.e.} $\alpha_{ep} < 6$)~\cite{feynman1955}.
We combine this with the property of the electron-phonon coupling constant~$\alpha_{ep}^2\propto 1/\omega_{o}$, to describe the oxygen-isotope dependence of the polaron-mass as a function of $m^*$
\begin{eqnarray}\label{polaron3}
\frac{d\ln{m^*}}{d\ln{M_{}}}\simeq \frac{1}{2} - \frac{13+\sqrt{90 m^*/m-65}}{36 m^*/m}
\end{eqnarray}

\newpage

\begin{figure}[ht]
\begin{center}
\includegraphics[width=0.8\columnwidth]{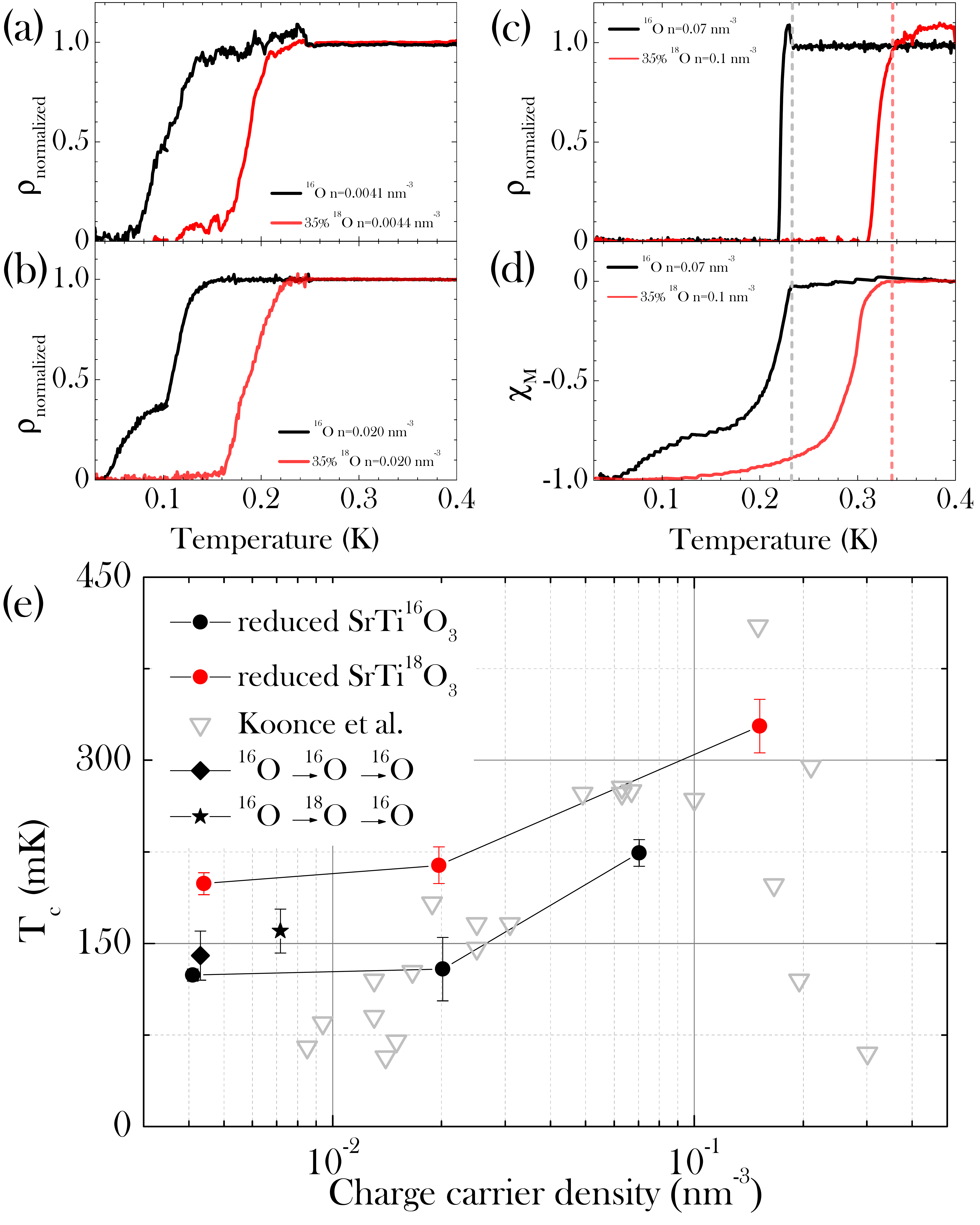}
\caption{\label{fig1} Normalized resistivity {\it vs.} temperature
at the superconducting transition of three different doping
levels: (a) $n=0.004$\,nm$^{-3}$ (b) $n=0.02$\,nm$^{-3}$ (c)
$n=0.07$\,nm$^{-3}$ (d) AC-Susceptibility showing the magnetic
transition to the superconducting state of the same sample as in
panel (c). (e) $T_c$ {\it vs.} charge carrier density. Full
symbols: experimental data of the present study for
SrTi$^{18}$O$_{3-y}$ (red) and SrTi$^{16}$O$_{3-y}$ (black). Grey
symbols: $T_c$ values reproduced from Ref.
~\onlinecite{koonce1967}. Black diamond and star refer to samples
in which $^{16}$O was back-substituted after isotope substitutions
(see {\it Methods}).}
\end{center}
\end{figure}

\begin{figure}[h]
\begin{center}
\includegraphics[width=0.75\columnwidth]{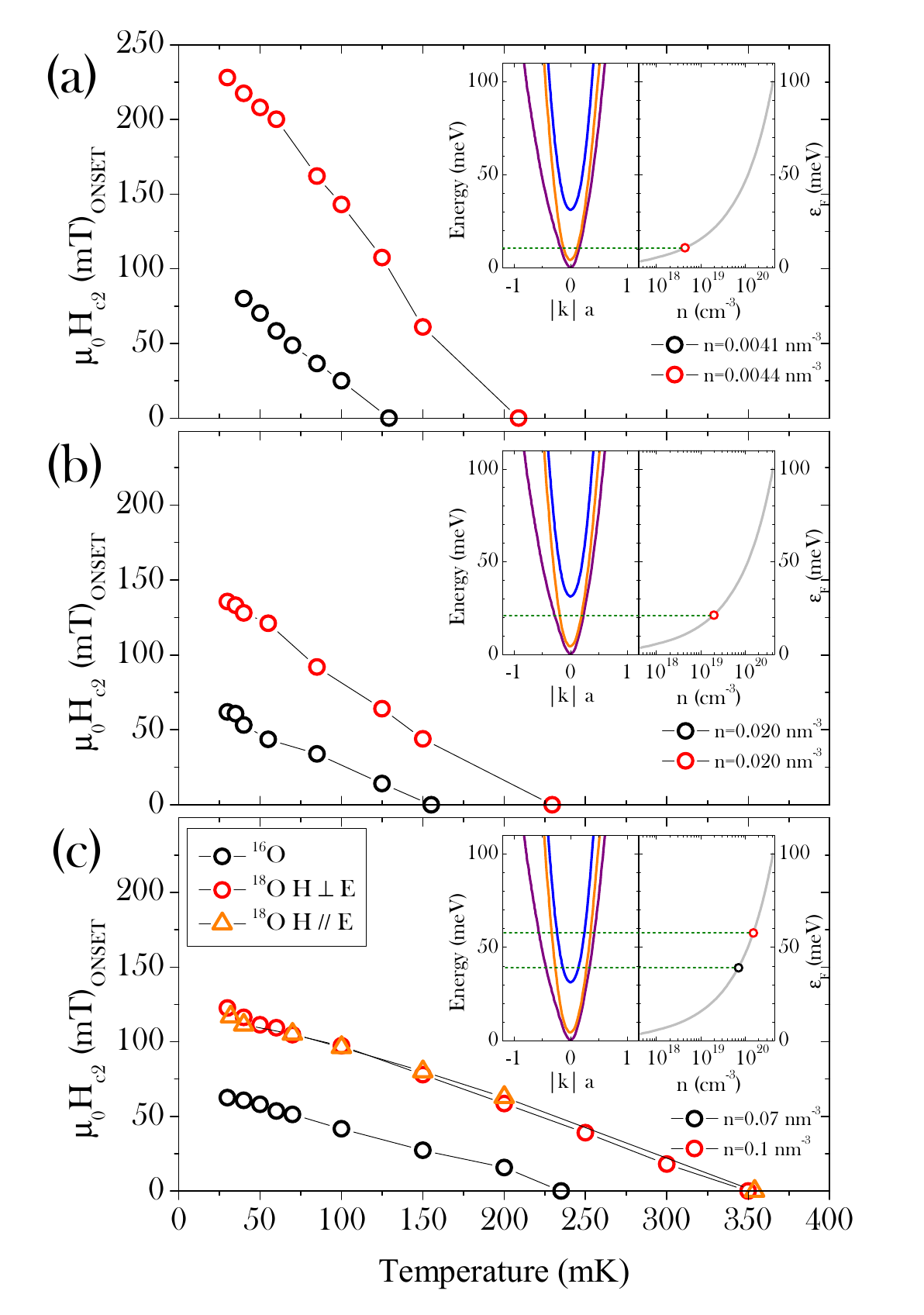}
\caption{\label{fig2} Upper critical field as a function of
temperature at various doping levels.
Black symbols:
SrTi$^{16}$O$_{3-y}$; red symbols:
 $^{18}$O-substituted SrTiO$_{3-y}$.
Panels (a), (b) and (c) refer to charge carrier densities $n\sim0.004, 0.02, 0.07$\,nm$^{-3}$, respectively.
 Open orange triangles and red circles in panel (c) indicate the H$_{c2}$ values measured with two mutually perpendicular directions of the applied magnetic field, both being perpendicular to the flowing current.
 Insets : detail of the conduction bands of SrTiO$_3$ and Fermi energy at each doping level [\onlinecite{vandermarel2011}].}
\end{center}
\end{figure}

\begin{figure}[hhh]
\begin{center}
\includegraphics[width=0.75\columnwidth]{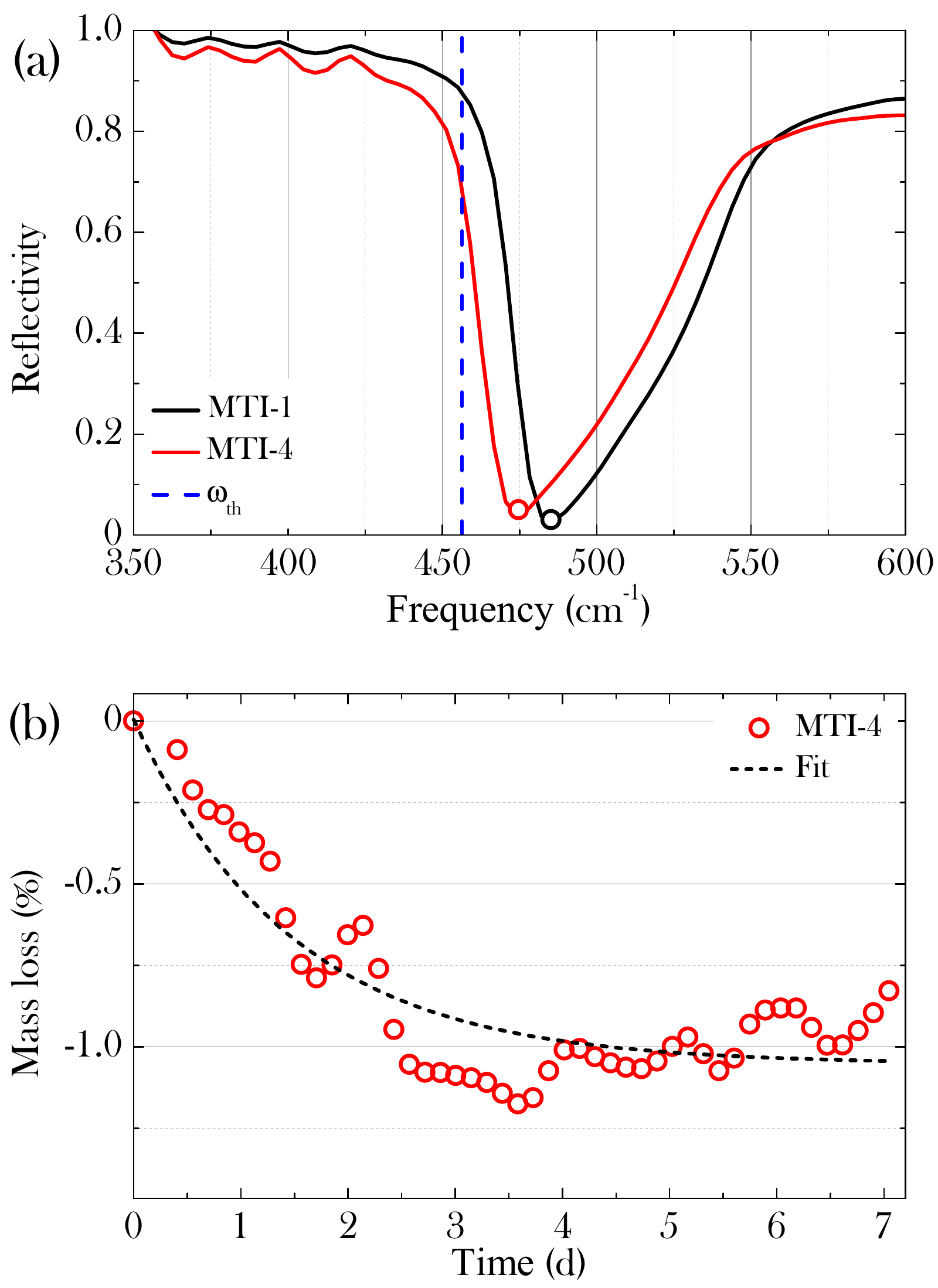}
\caption{\label{fig3} (a)Shift of the phonon mode in the IR
spectrum, due to isotope substitution. The vertical dashed line
correspond to the expected frequency shift in case of complete
isotope substitution. (b) Mass loss due to reverse substitution of
$^{16}$O for $^{18}$O as measured by thermogravimetry. The dashed
line is a fit to an asymptotic exponential.}
\end{center}
\end{figure}

\begin{figure}[t!]
\begin{center}
\includegraphics[width=0.75\columnwidth]{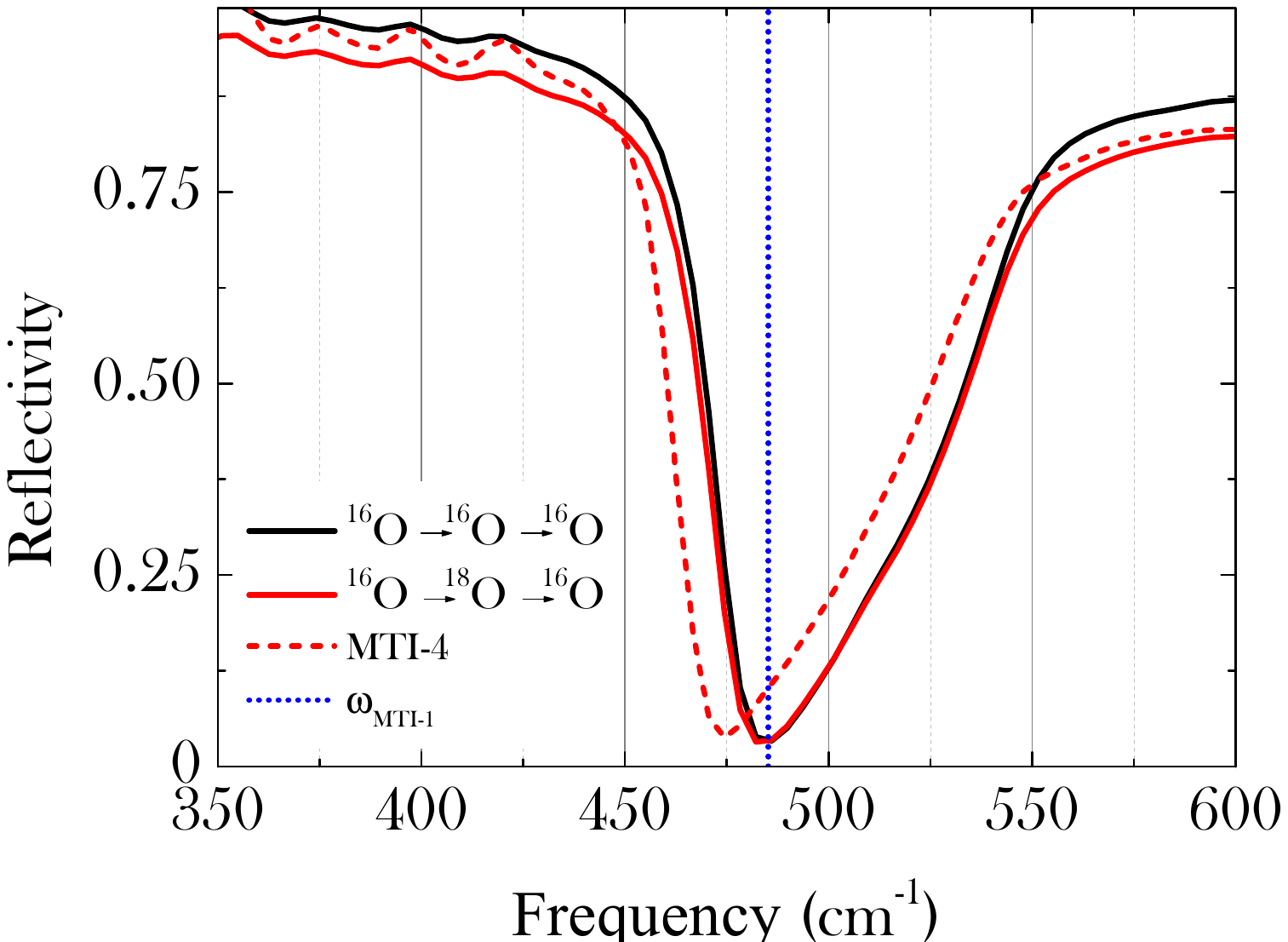}
\caption{\label{fig4} FIR spectroscopy of the two back-substituted samples (red and black lines). Red, respectively blue, dotted lines corresponding to the $^{18}$O substituted sample MTI-4, respectively to the phonon position of the pristine sample, are added as a reference for the phonon shift.}
\end{center}
\end{figure}


\newpage
\begin{thebibliography}{10}

\bibitem{mueller1979}
M\"uller, K.A. and Burkard, H. SrTiO$_3$: an intrinsic quantum paraelectric below 4 K. {\it Phys. Rev. B}, {\bf 19,} 3593-3602 (1979).

\bibitem{itoh1999}
Itoh, M. {\it et al.} Ferroelectricity induced by oxygen isotope exchange in strontium titanate perovskite. {\it Phys. Rev. Lett.} {\bf 82,} 3540-3543 (1999)

\bibitem{rowley2014}
Rowley, S. {\it et al.} Ferroelectric quantum criticality. {\it Nat. Phys.} {\bf 10,} 367-372 (2014).

\bibitem{eagles1969}
Eagles, D. M. Possible pairing without superconductivity at low carrier concentrations in bulk and thin-film superconducting semiconductors. {\it Phys. Rev.} {\bf 186,} 456-463 (1969).

\bibitem{calvani1993}
Calvani, P. {\it et al.} Observation of a midinfrared band in SrTiO$_{3-y}$. {\it Phys. Rev. B} {\bf 47,} 8917-8922 (1993).

\bibitem{gervais1993}
Gervais, F., Servoin, J.L., Baratoff, A., Bednorz,  J. G. and Binnig,  G. Temperature dependence of plasmons in Nb-doped ${\mathrm{SrTiO}}_{3}$. {\it
Phys. Rev. B} {\bf 47,} 8187-8194 (1993).

\bibitem{vanmechelen2008}
van Mechelen, J.L.M {\it et al.} Electron-phonon interaction and charge carrier mass enhancement in SrTiO$_3$. {\it Phys. Rev. Lett.} {\bf 100,} 226403 (2008).

\bibitem{meevasana2010}
Meevasana, W. {\it et al.} Strong energy-momentum dispersion of phonon-dressed carriers in the lightly doped band insulator SrTiO$_3$. {\it New J. Phys.} {\bf 12,} 023004 (2010).

\bibitem{vandermarel2011}
van~der Marel, D., van Mechelen, J.L.M. and Mazin, I. I. Common fermi-liquid origin of ${T}^{2}$ resistivity and superconductivity in $n$-type SrTiO$_3$. {\it Phys. Rev. B} {\bf 84,} 25111 (2011)

\bibitem{boschker2015}
Boschker, H., Richter, C., Fillis-Tsirakis, E., Schneider, C.W. and Mannhart, J. Electron-phonon coupling and the superconducting phase diagram of the
  LaAlO$_3$/SrTiO$_3$ interface. {\it Sci. Rep.}, {\bf 5,} 12309 (2015).

\bibitem{chen2015}
Chen, C., Avila, J., Frantzeskakis, E., Levy, A. and Asensio,M.C. Observation of a two-dimensional liquid of Fr\"ohlich polarons at the bare SrTiO$_3$ surface. {\it Nat. Commun.} {\bf 6,} 8585 (2015).

\bibitem{wang2016}
Wang, Z. {\it et al.} Tailoring the nature and strength of electron-phonon interactions in the SrTiO$_3$ (001) 2d electron liquid. {\it Nat. Mater.} {\bf 15,} 835-839 (2016).

\bibitem{cancellieri2016}
Cancellieri, C. {\it et al.} Polaronic metal state at the LaAlO$_3$/SrTiO$_3$ interface. {\it Nat. Commun.} {\bf 7,} 10386 (2016).

\bibitem{pfeiffer1969}
Pfeiffer, E. and Schooley, J. Superconducting transition temperatures of Nb-doped SrTiO$_3$. {\it Phys. Lett. A} {\bf 29,} 589-590 (1969).

\bibitem{suzuki1996}
Suzuki, H. {\it et al.} Superconductivity in single-crystalline Sr$_{1-x}$La$_x$TiO$_3$.  {\it J. Phys. Soc. Jpn}  {\bf 65,} 1529-1532 (1996).

\bibitem{schooley1964}
Schooley, J.F., Hosler, W.R. and Cohen, M.L. Superconductivity in semiconducting SrTiO$_3$. {\it Phys. Rev. Lett.} {\bf 12,} 474-475 (1964).

\bibitem{koonce1967}
Koonce, C.S., Cohen, M.L., Schooley, J.F., Hosler,  W.R. and Pfeiffer, E.R. Superconducting transition temperatures of semiconducting SrTiO$_3$. {\it Phys. Rev.} {\bf 163,} 380-390 (1967).

\bibitem{bednorz1988}
Bednorz, J.G. and M\"uller, K.A. Perovskite-type oxides - the new approach to high-${T}_{c}$ superconductivity. {\it Rev. Mod. Phys.} {\bf 60,} 585-600 (1988).

\bibitem{lin2014}
Lin, X. {\it et al.} Critical doping for the onset of a two-band superconducting ground state in SrTiO$_{3-\delta}$. {\it Phys. Rev. Lett.} {\bf 112,} 207002 (2014).

\bibitem{reyren2007}
Reyren, N. {\it et al.} Superconducting interfaces between insulating oxides. {\it Science} {\bf 317,} 1196-1199 (2007).

\bibitem{caviglia2008}
Caviglia, A.D. {\it et al.} Electric field control of the LaAlO$_3$/SrTiO$_3$ interface ground state. {\it Nature} {\bf 456} 624-627 (2008).

\bibitem{cohen1964}
Cohen, M.L. Superconductivity in many-valley semiconductors and in semimetals. {\it Phys. Rev.} {\bf 134,} A511 (1964).

\bibitem{appel1969}
Appel, J. Soft-mode superconductivity in SrTiO$_{3-x}$. {\it Phys. Rev.} {\bf 180,} 508-516 (1969).

\bibitem{zinamon1970}
Zinamon, Z. Superconductivity by small polarons. {\it Phil. Mag.} {\bf 21,} 347-356 (1970).

\bibitem{jarlborg2000}
Jarlborg, T. Tuning of the electronic screening and electron-phonon coupling in doped SrTiO$_3$ and WO$_3$. {\it Phys. Rev. B} {\bf 61,} 9887-9890 (2000).

\bibitem{klimin2012}
Klimin, S.N., Tempere, J., van der Marel, D. and Devreese, J.T. Microscopic mechanisms for the fermi-liquid behavior of Nb-doped strontium titanate. {\it Phys. Rev. B} {\bf 86,} 045113 (2012).

\bibitem{gorkov2016}
Gor'kov, L.P. Phonon mechanism in the most dilute superconductor ${n}$-type SrTiO$_{3}$. {\it Proc. Natl. Acad. Sci. of U.S.} {\bf 113,} 4646-4651 (2016).

\bibitem{edge2015}
Edge, J.M., Kedem, Y., Aschauer, U., Spaldin, N-.A. and Balatsky, A.V. Quantum critical origin of the superconducting dome in SrTiO$_3$. {\it Phys. Rev. Lett.} {\bf 115,} 247002 (2015).

\bibitem{kedem2016}
Kedem, Y., Zhu, J.-X. and Balatsky, A.V. Unusual superconducting isotope effect in the presence of a quantum criticality. {\it Phys. Rev. B} {\bf 93,} 184507 (2016).

\bibitem{jaccard2010}
Jaccard, D. and Sengupta, K. Multiprobe experiments under high pressure: resistivity, magnetic susceptibility, heat capacity, and thermopower measurements around 5 GPa. {\it Rev. Sci. Instrum.} {\bf 81,} 043908 (2010).

\bibitem{fowler1967}
Fowler, R.D., Lindsay, J.D.G., White, R.W., Hill, H.H. and Matthias, B.T. Positive isotope effect on the superconducting transition temperature of $\ensuremath{\alpha}$-uranium. {\it Phys. Rev. Lett.} {\bf 19,} 892-895 (1967).

\bibitem{bornemann1991}
Bornemann,H., Morris, D. and Liu, H. Negative oxygen isotope shift in Bi-2223 (Bi$_{1.6}$Pb$_{0.4}$Sr$_{2}$Ca$_{2}$Cu$_{3}$O$_{10}$) with $T_c$ = 108 K. {\it Physica C} {\bf 182,} 132-136 (1991).

\bibitem{stritzker1972}
Stritzker, B. and Buckel, W. Superconductivity in the Palladium-Hydrogen and the Palladium-Deuterium Systems. {\it Z. Phys.} {\bf 257,} 1-8 (1972).

\bibitem{miller1974}
Miller, R.J. and Satterthwaite, C.B. Electronic model for the reverse isotope effect in superconducting Pd-H(D). {\it Phys. Rev. Lett.} {\bf 34,} 144-148 (1975).

\bibitem{shirage2009}
Shirage, P.M. {\it et al.} Inverse iron isotope effect on the transition temperature of the (Ba,K)Fe$_2$As$_2$ superconductor. {\it Phys. Rev. Lett.} {\bf 103,}257003 (2009).

\bibitem{liu2009}
Liu, R.H. {\it et al.} A large iron isotope effect in SmFeAsO$_{1-x}$F$_{x}$ and Ba$_{1-x}$K$_{x}$Fe$_{2}$As$_{2}$. {\it Nature} {\bf 459,} 64-67 (2009).

\bibitem{schaeffer2015}
Schaeffer, A.M., Temple,  S.R., Bishop, J.K. and Deemyad, S. High-pressure superconducting phase diagram of $^6$Li: isotope effects in dense lithium. {\it Proc. Natl. Acad. Sci.  USA} {\bf 112,} 60-64 (2015).

\bibitem{binnig1980}
Binnig, G., Baratoff, A., Hoenig, H.E. and Bednorz, J.G. Two-band superconductivity in Nb-Doped SrTiO$_3$. {\it Phys. Rev. Lett.} {\bf 45,} 1352-1355 {1980}.


\bibitem{devreese2010}
Devreese, J.T., Klimin, S.N., van Mechelen, J.L.M. and van der Marel, D. Many-body large polaron optical conductivity in SrTi$_{1-x}$Nb$_x$O$_3$. {\it
Phys. Rev. B} {\bf 81,} 125119, (2010).

\bibitem{alexandrov1992}
Alexandrov, A.S. Transition from fermi liquid to charged bose liquid: a possible explanation of the isotope shift in high-$T_c$ oxides. {\it Phys. Rev. B} {\bf 46,} 14932 (1992).

\bibitem{feynman1955}
Feynman, R.P. Slow electrons in a polar crystal. {\it Phys. Rev.} {\bf 97,} 660-665 (1955).

\bibitem{khasanov2008}
Khasanov, R. {\it et al.} Oxygen isotope effects on the superconducting transition and magnetic states within the phase diagram of Y$_{1-x}$Pr$_{x}$Ba$_{2}$Cu$_{3}$O$_7$. {\it Phys. Rev. Lett.} {\bf 101} 077001 (2008).

\bibitem{bussmann2012}
Bussmann-Holder, A. and Keller, H.  Isotope and multiband effects in layered superconductors. {\it J. Phys. Conden. Matter} {\bf 24,} 233201 (2012).

\bibitem{hlinka2006}
Hlinka, J., Petzelt, J., Kamba, S., Noujni, D. and Ostapchuk, T. Infrared dielectric response of relaxor ferroelectrics. {\it Phase Transitions} {\bf 79,} 41-78 (2006).

\bibitem{spinelli2010}
Spinelli, A., Torija, M.A., Liu, C., Jan, C. and Leighton, C. Electronic transport in doped SrTiO$_3$: conduction mechanisms and potential applications. {\it Phys. Rev. B} {\bf 81,} 155110 (2010).

\end{thebibliography}
\end{document}